\documentclass[12pt]{article} 
\usepackage{fullpage} 
\usepackage[]{graphicx}

\newcommand{\ket}[1]{|#1 \rangle}

\begin{document}

\title{Generalised Hong-Ou-Mandel Experiments \\ with Bosons and Fermions}

\author{Yuan Liang Lim and Almut Beige \\[0.3cm]
{\small \em Blackett Laboratory, Imperial College London, Prince Consort Road,} \\[-0.1cm]
{\small \em London SW7 2BZ, United Kingdom}} 

\date{\today}

\maketitle 

\begin{abstract}
The Hong-Ou-Mandel (HOM) dip plays an important role in recent linear optics experiments. It is crucial for quantum computing  with photons and can be used to characterise the quality of single photon sources and linear optics setups. In this paper, we consider generalised HOM experiments with $N$ bosons or fermions passing simultaneously through a symmetric Bell multiport beam splitter. It is shown that for even numbers of bosons, the HOM dip occurs naturally in the coincidence detection in the output ports. In contrast, fermions always leave the setup separately exhibiting perfect coincidence detection. Our results can be used to verify or employ the quantum statistics of particles experimentally. 
\end{abstract}

\section{Introduction} \label{Introduction}

The two-photon Hong-Ou-Mandel (HOM) dip has been demonstrated first in 1987 \cite{Hong87}. In their experiment,  Hong, Ou and Mandel sent two identical photons simultaneously through the separate input ports of a $50:50$ beam splitter. Each output port contained a photon detector. Surprisingly, no coincidence detections within the temporal coherence length of the photons, i.e.~no simultaneous clicks in both detectors, were recorded. Crucial for the observation of this effect was  the identicalness of the pure quantum states of the input photons, which differed only in the directions of their wave vectors. This allowed the photons to become indistinguishable and to interfere within the setup. The detectors could not resolve the origin of each observed photon.

Due to the nature of this experiment, the HOM dip was soon employed for quantum mechanical tests of local realism \cite{Shih88} and for the generation of postselected entanglement between two photons \cite{Kwiat95}. Linear optics Bell measurements on photon pairs rely intrinsically on the HOM dip \cite{Ou,Braunstein95,Mattle96}, which has also been a building block for the implementation of linear optics gates for quantum information processing with photonic qubits \cite{Knill01}. Shor's factorisation algorithm \cite{Shor94}, for example, relies on multiple path interference to achieve massive parallelism \cite{Ou99} and multiphoton interference has to play a crucial role in any implementation of this algorithm using linear optics.  

Since it requires temporal and spatial mode-matched photons, observing the HOM dip for two photons is also a good test of their indistinguishability. HOM interference has been applied to characterise recently introduced sources for the generation of single photons on demand by testing the identicalness of successively generated photons \cite{Fattal04,Legero04,more}. Another interesting test based on the HOM dip has been studied by Bose and Home, who showed that it can reveal whether the statistics of two identical particles passing through a $50: 50$ beam splitter is fermionic or bosonic \cite{Bose02}. 

Motivated by the variety of possible applications of the two-photon HOM dip, this paper investigates generalised HOM experiments. We consider a straightforward generalisation of the scattering of two particles through a $50:50$ beam splitter, namely the scattering of $N$ particles through a symmetric $N \times N$ Bell multiport beam splitter. While numerous studies on $N$ photon interference in the {\em constructive} sense, i.e.~resulting in the enhancement of a certain photon detection syndrome, have been made (see e.g.~Refs.~\cite{Ou99}), not much attention has been paid to multiple path interference in the {\em destructive} sense. Recently Walborn {\em et al.} studied so-called multimode HOM effects for photon pairs with several inner degrees of freedom, including the spatial and the polarisation degrees of freedom \cite{Walborn03,Walborn04}. A notable example for destructive HOM interference has been given by Campos \cite{Campos00}, who studied certain triple coincidences in the output ports of an asymmetric $3 \times 3$ multiport beam splitter, which is also known as a tritter. 

\begin{figure}
\begin{minipage}{\columnwidth}
\begin{center}
\resizebox{\columnwidth}{!}{\rotatebox{0}{\includegraphics{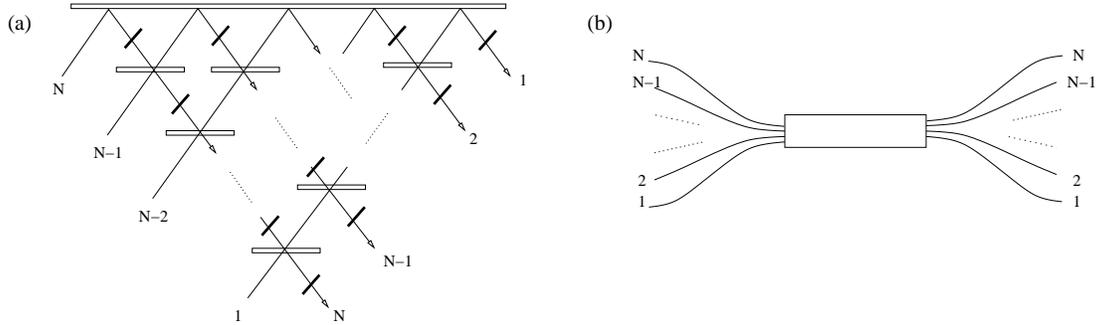}}}
\vspace*{-1cm} 
\end{center}
\caption{(a) Pyramidal construction of an $N \times N$ Bell multiport beam splitters for photons using a network of beam splitters and phase plates. (b) Alternatively, an $N \times N$ multiport can be constructed out of $N$ fibres spliced together.} \label{scheme}
\end{minipage}
\end{figure}

We consider {\em bosons} as well as the simultaneous scattering of {\em fermions}. The difference between both classes of particles is most elegantly summarised in the following commutation rules. While the annihilation and creation operators $a_i$ and $a_i^\dagger$ for a boson in mode $i$ obey the relation 
\begin{equation} \label{boson}
[a_i,a_j^\dagger] \equiv a_i a_j^\dagger - a_j^\dagger a_i = \delta_{ij} ~~~ {\rm and} ~~~ [a_i^\dagger,a_j^\dagger]=[a_i,a_j]=0 ~~ \forall ~ i, \, j 
\end{equation}
with $\delta_{ij}=0$ for $i \neq j$ and $\delta_{ii} = 1$, the annihilation and creation operators $a_i$ and $a_i^\dagger$ of fermionic particles obey the anticommutation relation
\begin{equation} \label{fermion}
\{a_i,a_j^\dagger\} \equiv a_i a_j^\dagger + a_j^\dagger a_i =\delta_{ij} ~~~ {\rm and} ~~~ \{a_i^\dagger,a_j^\dagger\}=\{a_i,a_j\}=0 ~~ \forall ~ i, \, j \, .
\end{equation} 
Here $i$ and $j$ refer to the inner degrees of freedom of the particles, like their respective path, polarisation, spin, frequency or energy.

Possible realisations of the $N \times N$ Bell multiport for photons, behaving like bosons, are shown in Figure \ref{scheme}. They may consist of a network of beam splitters and phase plates \cite{Reck94,Zukowski97} but can also be made by splicing $N$ optical fibers \cite{Pryde03}. Such spliced fibre constructions are commercially available and can include between $3$ and $30$ input and output ports.  The main feature of the symmetric Bell multiport is that a photon entering any of the input ports is equally likely redirected to any of the possible output ports. It can therefore be used to produce higher dimensional EPR correlations \cite{Zukowski97} and for the generation of so-called NOON states with special applications in lithography \cite{Pryde03,Kok02,Fiurasek02,Xubo02}. Moreover, the $N \times N$ Bell multiport can be used to prepare a great variety of multiphoton entangled states \cite{Lim04,Shi05} and to maximise the success probability when teleporting photonic qubits \cite{Knill01}. 

Furthermore, Bell multiport beam splitters exist for a wide variety of fermions and bosons. For example, multiports for bosonic or fermionic atoms can consist of a network of electrode wave guide beam splitters on an atom chip \cite{Cassettari00}. Multiports for electrons, who behave like fermions, can be realised by fabricating a network of quantum point contacts \cite{Samuelsson04}, split gates \cite{Henny99} or semiconductor multiterminal nanostructures \cite{Liu98,Oliver99} acting as two-electron beam splitters and can be used, for example, to detect two-electron Bell states \cite{Burkard00}. Specially doped optical fibres have recently been introduced in the literature and are expected to constitute beam splitters for ``fermion-like" photons \cite{Franson04}.

As in the original HOM experiment \cite{Hong87}, we assume in the following, that a particle detector is placed in each output port of  the scattering beam splitter array. The incoming particles should enter the different input ports more or less simultaneously and such that there is one particle per input port. Moreover, we assume that the particles are identical. It is shown, that it is impossible to observe a particle in each output port for even numbers $N$ of bosons. We denote this effect of zero coincidence detection as the {\em generalised HOM dip}. It is also shown that fermions always leave the setup separately exhibiting perfect coincidence detection. Since the interference behaviour of both types of particles is very different, the Bell multiport can be used to reveal their quantum statistics. 

This paper is organised as follows. In Section \ref{scatter} we introduce the theoretical description of particle scattering through a symmetric Bell multiport. Section \ref{two} describes the scattering of two particles through a $50 : 50$ beam splitter as an example. In Section \ref{many}, we derive the condition for the generalised HOM dip for bosons and analyse the scattering of fermions through the same setup for comparison. Finally we conclude our results in Section \ref{conclusions}.

\section{Scattering through a Bell multiport beam splitter} \label{scatter}

Let us first introduce the notation for the description of the scattering of $N$ particles through a passive setup. We proceed in close analogy to the scattering of photons through a linear optics network \cite{Zukowski97,Lim04}. Suppose each input port $i$ is entered by a particle with creation operator $a_i^\dagger$. Then the input state of the system equals 
\begin{eqnarray} \label{in}
|\phi_{\rm in} \rangle &=& \prod_{i=1}^N  a_i^\dagger  \, |0 \rangle \, ,
\end{eqnarray}
where $|0 \rangle$ is the vacuum state with no particles in the setup. Moreover, $S$ denotes the unitary scattering matrix, which connects the input state to its output state  
\begin{eqnarray} \label{fin}
|\phi_{\rm out} \rangle &=& S \, |\phi_{\rm in} \rangle \, .
\end{eqnarray}
Using Eq.~(\ref{in}) and the relation $S^\dagger S = 1$ then yields
\begin{eqnarray} \label{fin2}
|\phi_{\rm out} \rangle &=& S \,  a_1^\dagger \, S^\dagger S \,  a_2^\dagger  \cdot \, . \, . \, . \, \cdot S^\dagger S \, 
a_N^\dagger  \, S^\dagger S  \, |0 \rangle = \prod_{i=1}^N \, \, S \, a_i^\dagger \, S^\dagger \, \, |0 \rangle \, ,
\end{eqnarray}
since $S |0 \rangle =  |0 \rangle$ for the considered passive setup. As long as no particles enter the system, it remains in its vacuum state.

The Bell multiport beam splitter directs each incoming particle with equal probability to all output ports. To describe this, we introduce the matrix elements $U_{ji}$, which give the amplitude for the distribution of a particle from input port $i$ to output port $j$. Especially, for an $N \times N$ Bell multiport one has \cite{Zukowski97}
\begin{eqnarray}\label{fourier}
U_{ji} &=& {\textstyle{1 \over \sqrt{N}} \, \omega_N^{(j-1)(i-1)}} \, ,
\end{eqnarray} 
where $\omega_N$ denotes the $N$-th root of unity
\begin{equation} \label{root}
\omega_N \equiv {\rm e}^{2{\rm i} \pi /N }  \, .
\end{equation} 
The corresponding matrix $U$ is also known as the discrete Fourier transform matrix and has been widely considered in the literature \cite{Zukowski97,Pryde03,Lim04,Torma95,Kok01}. Proceeding as in Section II.D of Ref.~\cite{Zukowski97}, it can easily be verified that $U$ is unitary as well as symmetric. 

If $b_j^\dagger$ denotes the creation operator for a single particle in output port $j$, the definition of the scattering matrix $U$ implies that
\begin{eqnarray} \label{tran}
S \, a_i^\dagger \, S^\dagger = \sum_j U_{ji} \, b_j^\dagger \, .
\end{eqnarray}
Inserting this into Eq.~(\ref{fin2}), we obtain
\begin{eqnarray} \label{output1}
|\phi_{\rm out} \rangle &=& \prod_{i=1}^N \, \Bigg( \, \sum_{j=1}^N \, U_{ji} \, b_j^\dagger \, \Bigg) \, |0 \rangle \, . 
\end{eqnarray}
This equation describes the independent redirection of the incoming particles to their potential output ports. Conservation of the norm of the state vector is provided by the unitarity of the transition matrix $U$. Note that up to now, we have not invoked any assumptions about the nature of the particles. The formalism in this section applies to bosons and fermions equally.

\section{HOM interference of two particles} \label{two}

\begin{figure}
\begin{minipage}{\columnwidth}
\begin{center}
\vspace*{0cm}
\resizebox{\columnwidth}{!}{\rotatebox{0}{\includegraphics{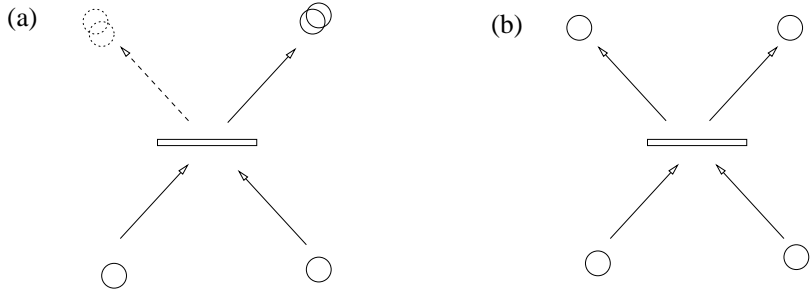}}}  
\end{center}
\vspace*{-1.5cm}
\caption{(a) HOM dip for two bosons scattering through a $50:50$ beam splitter. (b) Perfect coincidence in the output ports for fermion scattering.} \label{hom}
\end{minipage}
\end{figure}

Before analysing the general case, we motivate our discussion by considering two identical particles entering  the different input ports of a $50 : 50$ beam splitter. For $N=2$, the transition matrix (\ref{fourier}) becomes the Hadamard matrix
\begin{eqnarray}
U &=& {\textstyle {1 \over \sqrt{2}}} \left( \begin{array}{rr} 1 & 1 \\ 1 & -1 \end{array} \right)  
\end{eqnarray}
and the input state (\ref{in}) becomes $|\phi_{\rm in} \rangle=  a_1^\dagger a_2^\dagger \, |0 \rangle $. Note that local measurements on this input state cannot reveal any information about the bosonic or fermionic nature of the two particles. However, using Eq.~(\ref{output1}), we find that the beam splitter prepares the system in the state 
\begin{equation} \label{n2}
|\phi_{\rm out} \rangle = {\textstyle {1 \over 2}} \, \big(b_1^\dagger+b_2^\dagger \big) \big(b_1^\dagger-b_2^\dagger \big) \, |0 \rangle = {\textstyle {1 \over 2}} \left[ (b_1^\dagger)^2 - b_1^\dagger b_2^\dagger + b_2^\dagger b_1^\dagger - (b_2^\dagger)^2 \right] |0 \rangle \, .
\end{equation}
This state no longer contains any information about the origin of the particles, since each incoming one is equally likely                        
transferred to any of the two output ports. Passing through the setup, the input particles become indistinguishable by detection (see Figure \ref{hom}). Their quantum statistics can now be revealed using local measurements.

Bosons obey the commutation law (\ref{boson}). Using this, the output state (\ref{n2}) becomes
\begin{equation} 
|\phi_{\rm out} \rangle =  {\textstyle {1 \over 2}} \left[ (b_1^{\dagger})^2 - (b_2^{\dagger})^2 \right] |0 \rangle \, ,
\end{equation} 
which implies a zero-coincidence count rate at the output ports. The particles bunch together in the same output port and exhibit the famous HOM dip (see Figure \ref{hom}(a)). In contrast, fermions obey the anticommutation relation (\ref{fermion}) and their output state
\begin{equation} 
|\phi_{\rm out} \rangle = b_1^\dagger b_2^\dagger \, |0 \rangle
\end{equation}  
implies perfect particle coincidence. This means that the fermions always arrive in separate output ports and never bunch together (see Figure \ref{hom}(b)). A $50 : 50$ beam splitter can therefore indeed be used to distinguish bosons and fermions \cite{Bose02}. 

\section{Multiparticle HOM interference} \label{many}

We now consider the general case of $N$ particles passing through an $N \times N$ Bell multiport beam splitter. As in the $N=2$ case, the setup redirects each incoming particle with equal probability to any of the possible output ports, thereby erasing the information about the origin of each particle and making them indistinguishable by detection. For even numbers of bosons, this results in the generalised HOM dip and zero coincidence detection. In contrast, fermions leave the setup always separately, thus demonstrating maximum coincidence detection. Observing this extreme behaviour can be used, for example, to verify the quantum statistics of {\em many} particles experimentally. 

\subsection{Bosonic particles} \label{bos}

In order to derive the necessary condition for the appearance of the generalised HOM dip for even numbers of bosons, we calculate the output state (\ref{output1}) of the system under the condition of the collection of one particle per output port. Each term contributing to the projected conditional output state $|\phi_{\rm pro} \rangle$ can be characterised by a certain permutation, which maps the particles in the input ports $1, \, 2, \, ..., \, N$ to the output ports $1, \, 2, \, ..., \, N$. In the following, we denote any of the $N!$ permutations by $\sigma$ with $\sigma (i)$ being the $i$-th element of the list obtained when applying the permutation $\sigma$ onto the list $\{1,\, 2, \, ..., \, N\}$. Using this notation, $|\phi_{\rm pro} \rangle$ equals up to normalisation 
\begin{equation} \label{pro}
|\phi_{\rm pro} \rangle = \sum_{\sigma} \Bigg[ \prod_{i=1}^N U_{\sigma (i)i} \, b_{\sigma(i)}^{\dagger} \Bigg] \, |0 \rangle \, .
\end{equation} 
The norm of this state has been chosen such that 
\begin{equation} \label{suc}
P_{\rm coinc} = \| \, |\phi_{\rm pro} \rangle \, \|^2  
\end{equation}
is the probability to detect one particle per output port. It is therefore also the probability for observing coincidence counts in all $N$ detectors.

Up to now, the nature of the particles has not yet been taken into account. Using the commutation relation (\ref{boson}) for bosons, the conditional output state (\ref{pro}) becomes  
\begin{equation} \label{properm}
|\phi_{\rm pro} \rangle =  {\rm perm} \, U \cdot \prod_{i=1}^N b_i^{\dagger} \, \ket{0} 
\end{equation} 
with the permanent of the matrix $U$ defined as \cite{Scheel04,Horn85}
\begin{equation} \label{perm}
{\rm perm} \, U \equiv \sum_{\sigma} \prod_{i=1}^N U_{\sigma (i) \, i} \, .
\end{equation} 
The permanent of a matrix is superficially similar to the determinant. However, there exist hardly any mathematical theorems that can simplify the calculation of the permanent of an arbitrary matrix. 
 
To derive a condition for the impossibility of coincidence detections, we have to see when the probability (\ref{suc}) equals zero. Using Eq.~(\ref{properm}), we find 
\begin{equation} \label{almost}
P_{\rm coinc} = | \, {\rm perm} \, U \, |^2 \, .
\end{equation} 
The key to the following proof is to show that the transition matrix $U$ of the Bell multiport possesses a certain symmetry such that its permanent vanishes in certain cases. Suppose the matrix $U$ is multiplied by a diagonal matrix $\Lambda$ with matrix elements 
\begin{equation}
\Lambda_{jk}  \equiv \omega_N^{j-1} \, \delta_{jk} \, .
\end{equation} 
This generates a matrix $\Lambda U$ with 
\begin{eqnarray} \label{ma}
(\Lambda U)_{ji} = \sum_{k=1}^N \Lambda_{jk}U_{ki} = \Lambda_{jj}U_{ji} 
=  {\textstyle{1 \over \sqrt{N}}} \, \omega_N^{(j-1)i} \, .
\end{eqnarray} 
We now introduce the modulus function defined as ${\rm mod}_N (x)=j$, if $x-j$ is dividable by $N$ and $0<j<N$. Since  $\omega_N^ N = \omega_N^0 =1$, the matrix elements (\ref{ma}) can be expressed as
\begin{eqnarray}
(\Lambda U)_{ji}  =  {\textstyle{1 \over \sqrt{N}}} \, \omega_N^{(j-1)({\rm mod}_N (i)+1-1)} \, .
\end{eqnarray}
Note that the function $\tilde \sigma(i)={\rm mod}_N (i)+1$ maps each element of  the list $\{1,2,...N-1,N \}$ respectively to the list $\{2,3,...N,1\}$. A comparison with Eq.~(\ref{fourier}) therefore shows that 
\begin{equation}
(\Lambda U)_{ji} = U_{j \, \tilde \sigma(i)} \, .
\end{equation} 
In other words, the multiplication with $\Lambda$ amounts to nothing more than a cyclic permutation of the columns of the matrix $U$. Taking the cyclic permutation symmetry of the permanent of a matrix (see definition (\ref{perm})) into account, we obtain  
\begin{equation} \label{proof1}
{\rm perm} \, U ={\rm perm} \, (\Lambda U) \, .
\end{equation} 
However, we also have the relation
\begin{equation} \label{proof2}
{\rm perm} \, (\Lambda U) ={\rm perm} \, \Lambda \cdot {\rm perm} \, U 
\end{equation} 
with the permanent of the diagonal matrix $\Lambda$ given by
\begin{eqnarray} \label{relative} 
{\rm perm} \, \Lambda =  \prod_{k=1}^N \omega_N^{k-1} = \omega_N^{\sum_{k=1}^N k} = \omega_N^{N(N+1)/2} = {\rm e}^{ {\rm i} \pi (N+1)} = \left\{ \begin{array}{rl} 1 \, , & {\rm if}~N~{\rm is~odd} \, , \\  -1  \, , & {\rm if}~N~{\rm is~even} \, . \end{array} \right.
\end{eqnarray}  
For $N$ being even, a comparison of Eqs.~(\ref{proof1}) - (\ref{relative}) reveals that 
\begin{equation} \label{last}
{\rm perm} \, U = - {\rm perm} \, U  = 0 \, . 
\end{equation} 
As a consequence, Eq.~(\ref{almost}) implies that $P_{\rm coinc} =0$. Coincidence detection in all output ports of the setup is impossible for even numbers of bosons. This is not necessarily so, if the number of particles is odd. For example, for $N=3$ one can check that there is no HOM dip by calculating ${\rm perm} \, U$ explicitly. Campos showed that observing a HOM dip for $N=3$ is nevertheless possible with the help of a specially designed asymmetric multiport beam splitter \cite{Campos00}. 

\subsection{Fermionic particles}

Fermions scattering through a Bell multiport beam splitter show another extreme behaviour. Independent of the number $N$ of particles involved, they always leave the setup via different output ports, thereby guaranteeing perfect coincidence detection. As expected, particles obeying the quantum statistics of fermions cannot populate the same mode. 

Again, we assume that each input port is simultaneously entered by one particle and denote the creation operator of a fermion in output port $i$ by $b_i^\dagger$. Proceeding as in Section \ref{bos}, one finds again that the output state of the system under the condition of the collection of one particle per output port is given by Eq.~(\ref{pro}). To simplify this equation, we now introduce the sign function of a permutation with $\rm{sgn}(\sigma) = \pm 1$, depending on whether the permutation $\sigma$ is even or odd. An even (odd) permutation is one, that can be decomposed into an even (odd) number of integer interchanges. Using this notation and taking the anticommutator relation for fermions (\ref{fermion}) into account, we find 
\begin{equation}
|\phi_{\rm pro} \rangle = \sum_{\sigma}{\rm sgn}(\sigma) \Bigg( \prod_{i=1}^N U_{\sigma (i) \, i} \, b_i^{\dagger} \Bigg) \, |0 \rangle \, .
\end{equation} 
A closer look at this equation shows that the amplitude of this state relates to the determinant of the transformation matrix given by
\begin{equation}
{\rm det} \, U =\sum_{\sigma}{\rm sgn}(\sigma)  \prod_{i=1}^N U_{\sigma (i) \, i} \, .
\end{equation} 
Since $U$ is unitary, one has $|{\rm det} \, U| =1$ and therefore also, as Eq.~(\ref{suc}) shows, 
\begin{equation}
P_{\rm coinc} =  | \, {\rm det} \, U \, |^2 = 1 \, .
\end{equation} 
This means, that fermions always leave the system separately, i.e.~with one particle per output port. In the above, we only used the unitarity of the transition matrix $U$ but not it's concrete form.  Perfect coincidence detection therefore applies to any situation where fermions pass through an $N \times N$ multiport, i.e.~independent of its realisation.

\section{Conclusions} \label{conclusions}

We analysed a situation, where $N$ particles enter the $N$ different input ports of a symmetric Bell multiport beam splitter simultaneously. If these particles obey a fermionic quantum statistics, they always leave the setup independently with one particle per output port. This results in perfect coincidence detection, if detectors are placed in the output ports of the setup. In contrast to this, even numbers $N$ of bosons have been shown to never leave the setup with one particle per output port. This constitutes a generalisation of the two-photon HOM dip to the case of arbitrary even numbers $N$ of bosons. The generalised HOM dip is in general not observable when $N$ is odd. 

The proof exploits the cyclic symmetry of the setup. We related the coincidence detection in the output ports to the permanent or the determinant of the transition matrix $U$ describing the multiport, depending on the bosonic or fermionic nature of the scattered particles. Although the definition of the permanent of a matrix resembles that of the determinant, there exist almost no theorems to simplify their calculation \cite{Scheel04,Horn85}. In fact, the computation of the permanent is an NP-complete problem and much harder than the calculation of a determinant, which is only a P problem in complexity. Experimental setups involving the scattering of bosons through a  multiport therefore have important applications in quantum information processing.

For example, part of the linear optics quantum computing scheme by Knill, Laflamme and Milburn \cite{Knill01} is based on photon scattering through a Bell multiport beam splitter. In contrast to this, the scattering of non-interacting fermions through the same corresponding circuit, can be efficiently simulated on a classical computer \cite{Terhal02,Knill01a}. Moreover, the quantum statistics of particles has been used for a variety of quantum information processing tasks such as entanglement concentration \cite{Paunkovic02} and entanglement transfer \cite{Omar02}. Completely new perspectives might open when using setups that can change the quantum behaviour of particles and convert, for example, photons into fermions \cite{Franson04}.

Finally, we remark that observing HOM interference of many particles is experimentally very robust. Our results can therefore also be used to verify the quantum statistics of particles experimentally as well as to characterise or align an experimental setup. Testing the predicted results does not require phase stability in the input or output ports nor detectors with maximum efficiency. The reason is that any phase factor that a particle accumulates in any of the input or output ports contributes at most to an overall phase factor of the output state $|\phi_{\rm out} \rangle$. However, the coincidence statistics is sensitive to the phase factors accumulated inside the multiport beam splitter as they affect the form of the transition matrix (\ref{fourier}). The case, where the input ports of the setup are not entered by perfect one-particle states but by mixtures containing also a vacuum component can be analysed, in principle, using the methods introduced by Berry {\em et al.} in Refs.~\cite{sanders,sanders2}.

\vspace*{1cm}
\noindent {\em Acknowledgement.}  
The authors thank  J. D. Franson for valuable discussions. Y. L. L. acknowledges the DSO National Laboratories in Singapore for funding as a PhD student and A. B. thanks the Royal Society and the GCHQ for funding as a James Ellis University Research Fellow. This work was supported in part by the European Union and the UK Engineering and Physical Sciences Research Council.

\end{document}